\documentclass[12pt]{article}
\usepackage[latin1]{inputenc}
\usepackage{slashed}
\usepackage{amsmath}
\usepackage{color,graphicx}
\usepackage{textcomp}
\usepackage{amssymb}
\usepackage{amsfonts}
\usepackage{indentfirst}
\usepackage{color}
\usepackage[top=2cm,bottom=2cm,left=3cm,right=2cm]{geometry}
\newcommand{\nin}{\noindent}
\newcommand{\be}{\begin{equation}}
\newcommand{\ee}{\end{equation}}
\newcommand{\bea}{\begin{eqnarray}}
\newcommand{\eea}{\end{eqnarray}}

\newcommand{\nn}{\nonumber\\}

\begin{document}

\begin{flushright}
KCL-PH-TH/2015-06
\end{flushright}

\begin{center} 
 
 {\Large{\bf Dynamical mechanism for ultra-light\\ scalar Dark Matter}}

\vspace{0.5cm}

{\bf J. Alexandre} 

\vspace{0.2cm}

King's College London, Department of Physics, WC2R 2LS, UK\\
{\small jean.alexandre@kcl.ac.uk}

\vspace{1cm}

{\bf Abstract}

\end{center}

Assuming a double-well bare potential for a self-interacting scalar field, with the Higgs vacuum expectation value, 
it is shown that non-perturbative quantum corrections naturally lead to ultra-light particles of mass $\simeq10^{-23}$eV,
if these non-perturbative effects occur at a time consistent with the Electroweak phase transition.
This mechanism could be relevant in the context of Bose Einstein Condensate studies for the description of
cold Dark Matter. Given the numerical consistency with the Electroweak transition, an interaction potential for Higgs and 
Dark Matter fields is proposed, 
where spontaneous symmetry breaking for the Higgs field leads to the generation of ultra-light particles, in addition to 
the usual Higgs mechanism. 
This model also naturally leads to extremely weak interactions between the Higgs and Dark Matter particles.

\vspace{1.5cm}

\section{Introduction}

This article aims at providing a dynamical mechanism from which ultra-light scalar particles arise naturally, 
and which therefore could be relevant in the context of Bose-Einstein Condensate (BEC) Dark Matter (DM), 
proposed initially in \cite{Sin} and reviewed in \cite{review}. Axions have also been proposed 
as ultra-light particles to contribute to DM, for which a recent study can be found in \cite{marsh1} and a recent review in \cite{marsh2}. 

Models based on BEC DM assume the existence of a scalar particle for DM, light enough for its Compton wave length to be of the order of
the size of a DM halo. As a consequence, these scalars are in a coherent state, and can be described by a BEC wave function. It is usually assumed that 
condensed particles are non-relativistic, in order to describe cold DM, and that they decouple from the Standard Model (SM)
at some point in the Early Universe.

The mechanism proposed here, based on a non-perturbative effect in quantum field theory, explains how such ultra-light particle ($m<<$eV) can arise in the dressed theory, 
starting from a bare theory which contains a typical SM mass scale $v$. Such an effect is possible if the bare theory has
several degenerate minima, as the usual double-well potential for a single scalar field: since the dressed potential is necessarily convex \cite{Symanzik}, 
quantum corrections must be strong enough to eliminate the concave part of the bare potential, which allows for an ultra-small ratio $m/v$.  

Exploring the possible relevance of this mechanism to BEC DM, we find that the picture is consistent with non-perturbative quantum effects 
occurring at the Electroweak phase transition, where these ultra-light scalars could appear. The stability of such a BEC halo is discussed, and the repulsive self-interactions
predicted by this mechanism, although tiny, are enough to reduce substantially the range of fluctuation wave vectors which could potentially lead to a collapse of the condensate, 
under gravitational forces. 
Finally, we propose common dynamics between the Higgs and the DM particles, which explain how spontaneous symmetry breaking 
for the Higgs field could 
imply the generation of ultra-light particles, in addition to the usual Higgs mechanism. The corresponding bare potential 
for Higgs/DM naturally leads to extremely weak interactions between the Higgs and DM fields, after the non-perturbative effect for DM 
has taken place.

\section{Causality and ultra-light particles}

We review here the features of the convex effective potential (= momentum-independent part of the one-particle irreducible generating functional), for a finite volume.
This 4-dimensional volume is interpreted as the volume corresponding to the particle horizon, in a cosmological context, after requiring that quantum corrections 
should affect causally related events only.

\subsection{Convexity of the effective potential}

The convexity of the effective potential for a scalar theory has been known for a long time,
and is a consequence of its definition in terms of a Legendre transform \cite{Haymaker}.
In the situation where the bare potential has several degenerate minima, convexity
is achieved non-perturbatively, and cannot be obtained by a naive loop expansion. 
The effective potential becomes convex 
between the two minima of the bare potential, as a result of the contribution of several non-trivial saddle points in the partition function \cite{Fujimoto}.
The construction of the convex effective potential has been shown explicitly in \cite{AT}. We review here the results, but the derivation is generalised in the Appendix
of the present article, 
where the real scalar field is coupled to a complex scalar field, in order to justify the results presented in section 4.\\
We start from the generic double-well bare potential
\be\label{barepot}
U_{bare}(\varphi)=\frac{\lambda}{24}(\varphi^2-v^2)^2~,
\ee
and define the partition function $Z$ for a finite space time volume $V$. The semi-classical calculation is done by taking 
into account both minima of the bare potential, in a saddle point approximation, for the calculation of $Z$. The dressed potential is then, 
for $|\varphi|\leq v$ and in the limit of large space time volume $Vv^4>>1$,
\be\label{dressedpot}
U_{dressed}(\varphi)=\frac{1}{2V}\left(\frac{\varphi}{v}\right)^2+\frac{1}{12V}\left(\frac{\varphi}{v}\right)^4+{\cal O}(\varphi^6)
~~~\mbox{for}~~|\varphi|\leq v~.
\ee
As expected, this potential is convex, and higher orders in $\varphi$ are also suppressed by the volume (see Figure 1). Outside the minima of the bare potential $|\varphi|>v$, 
quantum corrections are perturbative, provided $|\varphi|$ is not too close to the minimum $v$, for the ``inside'' and ``outside'' potentials to match.
Note that the dressed potential (\ref{dressedpot}) is universal in the sense that it depends on the bare vacuum expectation value (vev) $v$ only, and not on
the bare coupling $\lambda$.

\begin{figure}
\begin{center}
\includegraphics[width=10cm,height=8cm]{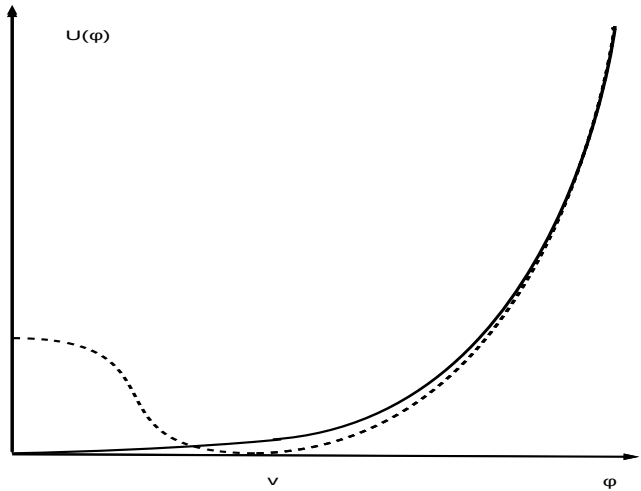}
\end{center}
\caption{\label{f1} Convexity of the effective potential as obtained in \cite{AT}, based on a semi-classical approximation to calculate the partition function.
The effective potential (continuous line) is non-perturbative for $\varphi^2<<v^2$, where the features of the bare potential (dotted line) 
are suppressed by the volume, as a consequence of the contribution of the two bare vacua. For $\varphi^2>>v^2$, only one bare vacuum dominates the partition function, and 
quantum corrections are perturbative. The region $\varphi^2\simeq v^2$ is sketched here to match the two previous regimes.}
\end{figure}

It has also been shown in \cite{AT} that the generalization to an $O(N)$-symmetric scalar theory gives, for $Vv^4>>1$, 
\be
U_{dressed}^{(N)}(\vec\varphi)=\frac{N}{2V}\left(\frac{\rho}{v}\right)^2+\frac{N^2}{4(N+2)V}\left(\frac{\rho}{v}\right)^4+{\cal O}(\rho^6)
~~~\mbox{for}~~|\rho|\leq v~,
\ee
where $\rho=\sqrt{\vec\varphi\cdot\vec\varphi}$, and the case of a complex scalar field is obtained for $N=2$. 

The non-perturbative mechanism described here can be understood with tunnelling effect\footnote{I would like to thank Arttu Rajantie for this comment}. 
For a finite volume, tunnelling between different vacua is allowed, such that the partition function is 
dominated by several saddle points, which leads to a convex effective potential. As a consequence, for finite volume, no spontaneous symmetry breaking occurs. It is only in the limit where
the volume goes to infinity that spontaneous symmetry breaking can take place, and the system chooses one vacuum among a continuous set of degenerate vacua.

Finally, a comment on the Higgs mechanism might be appropriate at this point. From the construction shown in \cite{AT}, one can understand that 
a convex dressed potential cannot be obtained if the scalar field if coupled to a gauge field. Indeed, in this situation, one needs to fix a gauge in order to 
define the path integral, such that the minima of the bare potential are not equivalent, and quantum fluctuations are built up from one scalar vacuum only,
defined by the choice of gauge. 

In what follows we will focus on the case $N=1$, since we are looking for a neutral scalar field to represent DM.

\subsection{Dynamical generation of ultra-light particles}

In the limit of infinite volume, the potential (\ref{dressedpot}) becomes flat between the minima of the bare potential. This
corresponds to the ``Maxwell construction'' in Statistical Mechanics, arising when two phases coexist in a system. In quantum field theory,
this flattening means that the true vacuum is a superposition of the different bare vacua \cite{Miransky}. 
This effect has motivated studies of inflation \cite{flat}, but in this note we will not take the limit of infinite volume. Instead
we consider the finite space time volume of causally related events $V=l_0^4$, where $l_0$ is the particle horizon
defined at a specific time $t_0$ to be determined below, and where quantum corrections drive  
the bare potential (\ref{barepot}) to its dressed form (\ref{dressedpot}).
One identifies then the dressed potential (\ref{dressedpot}) with the expression
\be
U_{dressed}(\varphi)\equiv\frac{1}{2}m^2\varphi^2+\frac{g}{24}\varphi^4~,
\ee
to find that the mass $m$ and the coupling constant $g$ are
\be\label{mg}
m=\frac{1}{vl_0^2}~~,~~~~~~g=\frac{2}{(vl_0)^4}~.
\ee
From the previous expression, the large volume condition leads to the inequalities
\be
\frac{1}{v}<<l_0<<\frac{1}{m}~,
\ee
showing that, necessarily, the dressed mass $m$ is small compared to the bare vev $v$.
We now consider the value $m\simeq10^{-23}$eV, which is typical in the context of BEC DM \cite{nointeraction, Lundgren}.
Also, in order to introduce a typical SM mass scale, we set $v$ to the Higgs vev $v=246$ GeV. We find then 
that the particle horizon is 
\be
l_0\simeq12~\mbox{cm}~, 
\ee
corresponding to the cosmological time 
\be
t_0=l/2\simeq2\times10^{-10}~\mbox{s}~,
\ee
for a radiation-dominated Universe. 
It is interesting to see that the choice of the Higgs vev for $v$ leads to a cosmological time consistent with 
the Electroweak phase transition, which suggests that the value $m\simeq10^{-23}$eV could indeed be related to this transition.
Finally, the large volume assumption is satisfied, since $vl_0\sim10^{17}$, which leads to 
\be\label{g}
g\sim10^{-68}~.
\ee
Although this coupling is extremely small, it contributes in a non-trivial way to the stability of the condensate, as explained below, because of the ultra-light scalar mass. 
We finally note here that detailed models for self-interacting BEC DM are studied in \cite{interaction}.

\section{Stability features of the condensate}

We present here two arguments towards the stability of the BEC DM halo, one related to the static halo, and the other related to its cosmological evolution.

\subsection{Gravitational collapse versus repulsive self interaction}

The Authors of \cite{Guth} consider small fluctuations of the non-relativistic
condensate, after linearising the corresponding equation of motion and studying the evolution in time of these fluctuations. 
They conclude that any repulsive interaction between scalars stabilises the condensate, but any attractive interaction 
induces instabilities though low wave vector fluctuations. In our present case, two effects compete: the attractive self-gravitation of the condensate
and the dynamical repulsive self-interaction controlled by $g>0$.\\
It is explained in \cite{Guth} that, if one takes into account gravity only, fluctuations with wave vector $k$ evolve in time as $\exp(\pm\gamma_0(k) t)$, with
\be
\gamma_0(k)=\frac{k}{2m}\sqrt{\frac{16\pi Gnm^3}{k^2}-k^2}~,
\ee
where $n$ is the number density of scalars of mass $m$ and $G$ is the Newton constant. It is easy to see that $\gamma_0$ vanishes for the critical wave vector $k^\star_0$ given by
\be
\frac{k^\star_0}{m}=\left(\frac{16\pi Gn}{m}\right)^{1/4}~.
\ee
$\bullet$ For $k>k^\star_0$, $\gamma_0$ is purely imaginary and fluctuations are purely oscillatory: the homogeneous configuration of the condensate is stable;\\
$\bullet$ For $k<k^\star_0$, $\gamma_0$ is real and the condensate is destabilised by low $k$-modes: the condensate collapses. \\
The typical mass density of our galaxy is 
\be\label{rhogalaxy}
\rho_{gal}\simeq1~\mbox{GeV~ cm}^{-3}~, 
\ee
and the number density of scalars is thus
\be\label{ngalaxy}
n\simeq \frac{1~ \mbox{GeV cm}^{-3}}{10^{-23} ~\mbox{eV}} \simeq10^{20}~ \mbox{eV}^3~, 
\ee
such that, together with $G\simeq 10^{-56}$ eV$^{-2}$, one finds
\be
\frac{k^\star_0}{m}\simeq10^{-3}~.
\ee
The fluctuation wave lengths $\simeq k^{-1}$ which could destabilise the condensate are therefore much larger than the BEC halo radius $\simeq m^{-1}$. 
If one takes into account the repulsive interaction controlled by the coupling constant (\ref{g}), $\gamma_0(k)$ is replaced by \cite{Guth}
\be
\gamma(k)=\frac{k}{2m}\sqrt{-\frac{gn}{2m}+\frac{16\pi Gnm^3}{k^2}-k^2}~,
\ee
and the critical wave vector $k^\star$ which satisfies $\gamma(k^\star)=0$ is now given by
\be
\left(\frac{k^\star}{m}\right)^2=\frac{gn}{4m^3}\left(-1+\sqrt{1+\frac{256\pi Gm^5}{g^2n}}\right)~.
\ee
Numerically, one finds
\be
\frac{k^\star}{m}\simeq10^{-16}<<\frac{k^\star_0}{m}~,
\ee
which reduces enormously the range of modes which potentially could destabilize the condensate. A more detailed analysis would be required, going beyond the linearised equation of motion for fluctuations,
to conclude on the definite stability of the condensate. But one can see here that the repulsive interaction predicted by the present model provides a huge improvement to the BEC halo stability.

\subsection{Time evolution}

The dressed potential (\ref{dressedpot}) holds as long as $|\varphi|\leq v$, but this condition remains valid 
for later times $t>t_0$, since the energy density decreases. 
More precisely, the equation of motion for the field is
\be\label{equamot}
\ddot\varphi+3H\dot\varphi+m^2\varphi=0~,
\ee
where the Hubble parameter is $H=(2t)^{-1}$ for the radiation-dominated era. We neglect here the back-reaction of the scalar's dynamics on gravity, which is 
assumed to be determined by radiation.
Since $H(t_0)\simeq4\times10^{-6}$eV$>>m$, one can initially neglect the mass term in eq.(\ref{equamot}),
which can be integrated once to give
\be
\dot\varphi=\dot\varphi_0\left(\frac{t_0}{t}\right)^{3/2}~,
\ee
where $\dot\varphi_0$ is a constant, corresponding to an initial condition.
The energy density is then, if one neglects the potential term $m^2\varphi^2/2$,
\be\label{rho}
\rho=\frac{1}{2}(\dot\varphi_0)^2\left(\frac{t_0}{t}\right)^3~,
\ee
and is proportional to the 6th inverse power of the scale factor $a(t)=t^{1/2}$, consistently with \cite{Lundgren}. 
The approximation (\ref{rho}) is not valid anymore around the 
time $t_1$, when the Hubble parameter $H(t_1)$ becomes equal to $m$. This time is 
\be
t_1\simeq 3\times10^7~\mbox{s}~,
\ee
which is still during the radiation-dominated
era. A more detailed analysis can be found in \cite{Lundgren}, showing that the energy density goes on decreasing, such that 
the amplitude of the field remains smaller than the bare vev $v$, and the potential (\ref{dressedpot}) 
remains appropriate for the description of 
non-interacting ultra-light particles.

\section{Higgs/DM dynamics}

We suggest here common dynamics to the Higgs and the ultra-light DM particles, given the above numerical coincidence. 
The motivation is to generate ultra-light particles simultaneously with the Higgs mechanism. As shown here, 
the non-vanishing vev of the Higgs field could also generate the double-well bare potential for the DM neutral field, 
which therefore would lead to ultra-light scalars, after non-perturbative quantum corrections.

For simplicity, we consider the Abelian Higgs model, and extend it with the neutral DM scalar $\varphi$
\be\label{AbelianHiggs}
{\cal L}=-\frac{1}{4}F_{\mu\nu}F^{\mu\nu}+D_\mu\phi(D^\mu\phi)^\star+\frac{1}{2}\partial_\mu\varphi\partial^\mu\varphi-U_{bare}(\phi\phi^\star,\varphi^2)~,
\ee
where $\phi$ is the complex Higgs field and $D_\mu$ is the covariant derivative. The potential is chosen as
\be\label{bareHiggs}
U_{bare}=\frac{g_H}{24}\left[\left(\phi\phi^\star- v^2\right)^2+\left(\phi\phi^\star-\varphi^2\right)^2\right]~,
\ee
which is minimised for the field values
\be
(\phi\phi^\star)_0~=\varphi_0^2= v^2~.
\ee
The derivation of the semi-classical effective potential for DM is given in the Appendix, which generalises the original derivation in \cite{AT} to the 
situation where the real scalar field is coupled to a complex scalar field. The result found in the Appendix can be obtained in a simpler way though,
as follows. Because the Higgs field is coupled to the gauge field, its vev $v$ is unique and is determined by spontaneous symmetry breaking. 
But this is no the case for the real field $\varphi$ which, for a given Higgs field configuration, sees the two bare minima $\pm \sqrt{\phi\phi^\star}$. 
The non-perturbative mechanism described in section 2 leads then to the DM effective potential 
\be
U^{eff}_{DM}=\frac{1}{2V}\frac{\varphi^2}{\phi\phi^\star}+\frac{1}{12V}\frac{\varphi^4}{(\phi\phi^\star)^2}+{\cal O}(\varphi^6)~~~\mbox{for}~\varphi^2<\phi\phi^\star~.
\ee
We then parametrise the Higgs field in the usual way
\be
\phi\equiv v+\xi+i\eta~,
\ee
where $\xi$ is the dynamical Higgs field and the Goldstone mode $\eta$ is absorbed as a longitudinal component of
the vector field, and is therefore ignored in what follows. An expansion of $\phi\phi^\star$ around $v^2$ to the second order in $\xi$, leads to
\be\label{UeffDM}
U^{eff}_{DM}(\varphi,\xi)=\frac{1}{2V}\left(\frac{\varphi}{v}\right)^2+\frac{1}{12V}\left(\frac{\varphi}{v}\right)^4
-\frac{1}{V}\frac{\varphi^2\xi}{v^3}+\frac{3}{2V}\frac{\varphi^2\xi^2}{v^4}+\cdots~,
\ee
where dots represents higher-order interactions terms. This result is consistent with the full derivation given in the Appendix, and describes the following Higgs-DM interactions: 
\begin{itemize} 

\item A decay process Higgs$\to$DM-DM which, given the ultra-small ratio $m/m_{Higgs}\simeq10^{-34}$, leads to relativistic DM particles, not contributing to the BEC.
These relativistic DM particles are predicted in BEC DM models \cite{review}, and consist in a small portion of DM particles. This cubic interaction also represents an annihilation of two DM
particles into a Higgs particle, which is kinematically possible with relativistic DM only; 

\item A scattering process Higgs-Higgs$\to$DM-DM or Higgs-DM$\to$Higgs-DM, which is repulsive and therefore stabilises further the DM halo, according to the analysis of section 3.1. 

\end{itemize}

\nin These two interactions are highly suppressed though, by coupling constants which are proportional to $(Vv^4)^{-1}\simeq10^{-68}$,
and therefore are not detectable. \\
This conclusion holds as long as the amplitude of $\varphi^2$ is smaller than $v^2$, where DM particles see the effective 
potential (\ref{UeffDM}). In a regime where the number density of DM particles is such that $\varphi^2>v^2$ though, the potential seen by 
these particles consist in perturbative corrections to the bare potential (\ref{bareHiggs}), such that the interaction DM/Higgs could be 
detected. The critical field value $\varphi^2=v^2$ characterising the transition between perturbative and non-perturbative effective potentials (see Figure 1) 
corresponds to an energy density of the order $U^{eff}_{DM}(v,0)\simeq V^{-1}$, and therefore to the number density of DM particles
\be
n_{crit}=\frac{U^{eff}_{DM}(v,0)}{m}=\frac{v}{l_0^2}= \frac{246~\mbox{GeV}}{(12~\mbox{cm})^2}\simeq10^{57}~\mbox{eV}^3~.
\ee
The ratio between $n_{crit}$ and the number density (\ref{ngalaxy}) of DM in the galaxy is huge
\be
\frac{n_{crit}}{n_{gal}}=10^{37}~,
\ee
but can be achieved during the quark-gluon plasma phase generated in heavy-ion collisions. Indeed, the critical energy density for a quark-gluon plasma to be created 
corresponds to the proton energy density
\be
\rho_{prot}\simeq 1~\mbox{GeV~fm}^{-3}~,
\ee
whose ratio with the galaxy energy density (\ref{rhogalaxy}) is
\be
\frac{\rho_{prot}}{\rho_{gal}}=10^{39}~.
\ee
An experimental signature for the present model through the study of quark-gluon plasma might therefore be possible, although much more work needs to be done in this direction.

\section{Conclusion}

The main point in this article consists in the identifications (\ref{mg}), which allow the hierarchy between the 
scales $m$ and $v$ in natural way, and
predict that ultra-light particles have repulsive self-interactions which improves the stability of the BEC DM. 
The consistency of the picture with the Electroweak transition shows a new avenue to explore: the link between 
the Higgs particle and Dark Matter. Both particles could see the same vev, but behave in a different way, according to 
their coupling to gauge fields: the Higgs field sees perturbative quantum corrections, whereas the additional scalar, blind to gauge fields, 
undergoes non-perturbative corrections and becomes ultra-light. 
The common dynamics between the Higgs and the DM fields, presented in section 4, is a first step in the direction 
of a potential unification of the two fundamental origins of mass in the Universe.

\section*{Appendix: Semi-classical derivation of the DM effective potential}

We generalise here the derivation given in \cite{AT} to the situation where the real scalar field is coupled to a complex scalar field 
which sees only one bare vacuum because of gauge fixing. Starting with the Abelian model (\ref{AbelianHiggs}), we focus on the scalar sector. 
As shown below, the main features of the mechanism are:
\begin{itemize}
\item Because of gauge fixing, the Higgs field sees only one vacuum, and the corresponding
semi-classical approximation doesn't modify the bare Higgs sector;
\item The real scalar field sees two bare vacua, since it is not coupled to the gauge field. Tunnelling effect - for finite volume - leads 
then to the non-perturbative mechanism which results in the suppression of the effective potential by the volume.
\end{itemize}
We follow here the usual steps of path integral quantisation, where the Euclidean partition function, functional of the sources $j,j^\star,k$, is 
\be
Z[j,j^\star,k]=\int{\cal D}[\phi,\phi^\star,\varphi]\exp\left(-\int_x\partial_\mu\phi\partial^\mu\phi^\star+\frac{1}{2}\partial_\mu\varphi\partial^\mu\varphi
+U_{bare}(\rho,\varphi)+j\phi+j^\star\phi^\star+k\varphi\right)~,
\ee
where $\rho=\sqrt{\phi\phi^\star}$ and 
\be
U_{bare}(\rho,\varphi)=\frac{g_H}{24}\left[\left(\rho^2- v^2\right)^2+\left(\rho^2-\varphi^2\right)^2\right]~.
\ee
We choose the (unique) Higgs vacuum to be in the real direction and disregard the Goldstone mode, such that $\phi^\star=\phi$ and the partition function reads
\be\label{Z}
Z[r,k]=\int{\cal D}[\phi,\varphi]\exp\left(-\int_x\partial_\mu\phi\partial^\mu\phi+\frac{1}{2}\partial_\mu\varphi\partial^\mu\varphi
+U_{bare}(\rho,\varphi)+2r\rho+k\varphi\right)~,
\ee
where $r=|j|$. Due to the double-well structure of the potential,
this partition function is dominated by two uniform saddle point configuration \footnote{This is true as long as the sources are smaller than critical sources,
or equivalently the classical fields (defined below) are smaller than the bare vacuum $v$, which is what we assume here. As explained in \cite{AT}, 
for classical fields larger than $v$ (or equivalently for large sources), the partition function is dominated by one configuration only, and the semi-classical 
approximation that we use below, to determine the effective potential, simply leads to the bare potential for these field values.} $(\rho_+,\phi_+)$ and $(\rho_-,\phi_-)$.
These saddle points are obtained after minimizing the action, for uniform sources, and are solution of the equations  
\be
\tilde\varphi_0(\tilde\varphi_0^2-\tilde\rho_0^2)+\tilde k=0~~~~\mbox{and}~~~~
\tilde\rho_0(2\tilde\rho_0^2-\tilde\varphi_0^2-1)+\tilde r=0~,
\ee
where the dimensionless fields are
\be
\tilde\rho_0\equiv\frac{\rho_0}{v}~,~~\tilde\varphi_0\equiv\frac{\varphi_0}{v}~,~~\tilde r\equiv\frac{12 r}{g_H v^3}~,~~\tilde k\equiv\frac{6 k}{g_H v^3}~,
\ee
and $\varphi_0=\varphi_\pm$, $\rho_0=\rho_\pm$. A Taylor expansion gives for the first saddle point
\bea
\tilde\rho_+&=&1-\tilde r/2-\tilde k/2-3\tilde r^2/8-3\tilde r\tilde k/4-5\tilde k^2/8-\tilde r^3/2-3\tilde r^2\tilde k/2-9\tilde r\tilde k^2/4-25\tilde k^3/16\nn
&&-105\tilde r^4/128-105\tilde r^3 \tilde k/32-441\tilde r^2\tilde k^2/64-69\tilde r\tilde k^3/8-637\tilde k^4/128+\cdots\nn
\tilde\varphi_+&=&1-\tilde r/2-\tilde k-3\tilde r^2/8-5\tilde r\tilde k/4-33\tilde k^2/3-\tilde r^3/2-9\tilde r^2\tilde k/4-75\tilde r\tilde k^2/16-4\tilde k^3\nn
&&-105\tilde r^4/128-147\tilde r^3 \tilde k/32-207\tilde r^2\tilde k^2/16-637\tilde r\tilde k^3/32-105\tilde k^4/8+\cdots
\eea
and for the second
\bea
\tilde\rho_-&=&1-\tilde r/2+\tilde k/2-3\tilde r^2/8+3\tilde r\tilde k/4-5\tilde k^2/8-\tilde r^3/2+3\tilde r^2\tilde k/2-9\tilde r\tilde k^2/4+25\tilde k^3/16\nn
&&-105\tilde r^4/128+105\tilde r^3 \tilde k/32-441\tilde r^2\tilde k^2/64+69\tilde r\tilde k^3/8-637\tilde k^4/128+\cdots\nn
\tilde\varphi_-&=&-1+\tilde r/2-\tilde k+3\tilde r^2/8-5\tilde r\tilde k/4+33\tilde k^2/3+\tilde r^3/2-9\tilde r^2\tilde k/4+75\tilde r\tilde k^2/16-4\tilde k^3\nn
&&+105\tilde r^4/128-147\tilde r^3 \tilde k/32+207\tilde r^2\tilde k^2/16-637\tilde r\tilde k^3/32+105\tilde k^4/8+\cdots
\eea
where dots represent higher orders in the sources.
The partition function is then determined by the saddle point approximation
\bea
Z[r,k]&\simeq&\frac{1}{2}\exp\left(V[U_{bare}(\rho_+,\varphi_+)+2r\rho_+ +k\varphi_+]\right)\nn
&&+\frac{1}{2}\exp\left(V[U_{bare}(\rho_-,\varphi_-)+2r\rho_- +k\varphi_-]\right)~,
\eea
where $V$ is the space time volume. A Taylor expansion in the sources gives
\bea
Z[\tilde r,\tilde k]&=&1+4A\tilde r+8A^2\tilde r^2+8A^2\tilde k^2+(32A^3/3)\tilde r^3+32A^3\tilde r\tilde k^2\nn
&&+(32A^4/3)\tilde r^4+64A^4\tilde r^2\tilde k^2+(32A^4/3)\tilde k^4+\cdots
\eea
where
\be
A\equiv\frac{g_HVv^4}{24}~,
\ee
and only the dominant terms in $A>>1$ are kept for the large volume approximation.
For a finite volume and constant field configurations, the functional derivatives become partial derivatives and the classical fields are given by 
\bea
\rho_c\equiv v\tilde\rho_c=\frac{1}{2Z}\left|\frac{\delta Z}{\delta r}\right|&\to&\tilde\rho_c=\frac{1}{4AZ}\left|\frac{\partial Z}{\partial\tilde r}\right|\nn
\varphi_c\equiv v\tilde\varphi_c=-\frac{1}{Z}\frac{\delta Z}{\delta k}&\to&\tilde\varphi_c=\frac{-1}{4AZ}\frac{\partial Z}{\partial\tilde k}
\eea
In the large volume approximation $A>>1$ we find the Taylor expansions
\bea\label{negative}
\tilde\rho_c&=&1-(1/2)\tilde r-(3/8)\tilde r^2-2A\tilde k^2-(1/2)\tilde r^3-2A\tilde r \tilde k^2\nn
&&-(105/128)\tilde r^4-(15A/4)\tilde r^2\tilde k^2-(32A^3/3)\tilde k^4+\cdots\nn
\tilde\varphi_c&=&-4A\tilde k+4A\tilde r\tilde k+2A\tilde r^2\tilde k+(64A^3/3)\tilde k^3\nn
&&+(5A/2)\tilde r^3\tilde k+(128A^3/3)\tilde r\tilde k^3+\cdots~,
\eea
where one can see that, whereas the complex classical amplitude $\rho_c$ oscillates around the vev $v$, the real scalar $\phi_c$ oscillates around 0 because of tunnelling effect. 
The latter relations can be inverted to obtain the sources as functions of the fluctuations $\tilde\xi_c=\tilde\rho_c-1$ and $\tilde\varphi_c$: 
\bea\label{tilderxi}
\tilde r&=&-2\tilde\xi_c-3\tilde\xi_c^2-\tilde\varphi_c^2/(4A)-\tilde\xi_c^3+3\tilde\xi_c\tilde\varphi_c^2/(4A)\nn
&&-3\tilde\xi_c^2\tilde\varphi_c^2/(2A)-\tilde\varphi_c^4/(12A)+\cdots\nn
\tilde k&=&\tilde\varphi_c/(4A)-\tilde\xi_c\tilde\varphi_c/(2A)+3\tilde\xi_c^2\tilde\varphi_c/(4A)+\tilde\varphi_c^3/(12A)\nn
&&-\tilde\xi_c^3\tilde\varphi_c/A-\tilde\xi_c\tilde\varphi_c^3/(3A)+\cdots
\eea
where only the dominant terms in $A>>1$ are considered. Note that, in the present semi-classical approximation, the fluctuations $\tilde\xi_c$ are negative (which can be see from eq.(\ref{negative})),
such that $\tilde r$ in eq.(\ref{tilderxi}) is positive, as expected, since it is the modulus of the source $\tilde j$.

The effective action for uniform fields 
\be
\Gamma[\xi_c,\varphi_c]=VU^{eff}(\xi_c,\varphi_c)~,
\ee
is defined as the Legendre transform of the connected graphs generating functional 
\be
W[r,k]\equiv -\ln Z[r,k]~,
\ee
and reads, for uniform classical fields,
\be
\Gamma[\xi_c,\varphi_c]=W[r,k]-V(2r\xi_c+k\varphi_c)~,
\ee
where the sources $r,k$ have to be understood as functions of the classical fields $\xi_c,\varphi_c$ through the relations (\ref{tilderxi}). 
From its definition, it is known that $\Gamma$ satisfies the two following equations
\bea
r=v\tilde r=\frac{1}{2}\left|\frac{\delta\Gamma}{\delta\xi_c}\right|&\to&\tilde r=\frac{1}{4A}\left|\frac{\partial\Gamma}{\partial\tilde\xi_c}\right|\nn
k=v\tilde k=-\frac{\delta\Gamma}{\delta\varphi_c}&\to&\tilde k=\frac{-1}{4A}\frac{\partial\Gamma}{\partial\tilde\varphi_c}~,
\eea
which, together with the expansions (\ref{tilderxi}), can be integrated to get
\bea
\Gamma&=&4A\tilde\xi_c^2+4A\tilde\xi_c^3+A\tilde\xi_c^4+\frac{1}{2}\tilde\varphi_c^2+\frac{1}{12}\tilde\varphi_c^4-\tilde\xi_c\tilde\varphi_c^2+\frac{3}{2}\tilde\rho_c^2\tilde\varphi_c^2+\cdots\nn
&=&\frac{g_HVv^4}{24}\left((1+\tilde\xi_c)^2-1\right)^2+\frac{1}{2}\tilde\varphi_c^2+\frac{1}{12}\tilde\varphi_c^4-\tilde\xi_c\tilde\varphi_c^2+\frac{3}{2}\tilde\rho_c^2\tilde\varphi_c^2+\cdots
\eea
The effective potential is finally obtained after dividing by the volume $V$
\be
U^{eff}=\frac{g_H}{24}\left((v+\xi_c)^2-v^2\right)^2+\frac{1}{V}\left[\frac{1}{2}\left(\frac{\varphi_c}{v}\right)^2+\frac{1}{12}\left(\frac{\varphi_c}{v}\right)^4-\frac{\xi_c\varphi_c^2}{v^3}
+\frac{3}{2}\frac{\xi_c^2\varphi_c^2}{v^4}\right]+\cdots
\ee
and contains two contributions:
\begin{itemize}
\item The part proportional to $g_H$, which is the usual bare Higgs potential, and has not been modified in this semi-classical approximation;
\item The part depending on $\varphi_c$ obtained in eq.(\ref{UeffDM}) with a simpler argument, which is suppressed by the volume. 
\end{itemize}

\end{document}